\documentclass[aps,prb,twocolumn,superscriptaddress,noshowpacs]{revtex4}

\usepackage{graphicx}
\usepackage{amsfonts,amsmath,amssymb,amsthm}

\def\be{\begin{equation}}
\def\ee{\end{equation}}
\def\ba{\begin{eqnarray}}
\def\ea{\end{eqnarray}}

\def\bc{\begin{center}}
\def\ec{\end{center}}

\begin{document}

\title{

Friedel oscillations as an origin of Mahan exciton$/$Fermi edge singularity phenomena in optical spectra of degenerated semiconductors}

\author{V.~V.~Solovyev}  

\email{vicsol@issp.ac.ru}

\affiliation{Institute of Solid State Physics, RAS, Chernogolovka
142432, Russia}

\author{I.~V.~Kukushkin}
\affiliation{Institute of Solid State Physics, RAS, Chernogolovka
142432, Russia}

\date{\today}

\begin{abstract}

We point out at the possible relationship between the two many-body phenomena: Friedel oscillations and Mahan exciton$/$Fermi edge singularity in a recombination process of a localized valence band hole with a degenerate electron system. It is qualitatively described that the oscillating electron density around the screened potential of a neutral acceptor can modify the wavefunction of the valence band hole and thus promote its recombination with electrons near Fermi-surface. The role of a finite hole mass for the net balance of the two hole energies: the kinetic energy and Coulomb attraction to the degenerate electron system, - is discussed and explicitly shown to be crucial for the occurrence of Mahan enhancement phenomena. 

\end{abstract}

\maketitle

The presence of a Fermi surface in degenerate electronic systems that separates the occupied electron states in a momentum space from the unoccupied ones is the ground for literally all physical features of the fermionic ensembles possessing a 'normal' ground state. Originating from the very fundamental concept of the Pauli exclusion principle, it corresponds to the invariant of the Fermi momentum $k_F$ in the electron momentum space. This invariant is dictated solely by the electron density and survives even after transforming the Fermi-gas of non-interacting electrons into the Fermi-liquid with an arbitrary strong interaction ~\cite{Luttinger}. 

Of all the phenomena arising thanks to a Fermi surface presence, we'd like to discuss in this communication and try to connect the two ones that are hardly related at a first glance: Friedel oscillations ~\cite{Friedel} and Mahan exciton$/$Fermi edge singularity ~\cite{Mahan} in optical absorption and emission spectra. 

The latter one originates from the pioneering theoretical analysis by Mahan considering the response of an electron sea to the event of creation or annihilation of either a core level or valence band hole that interacts with electrons via Coulomb attraction. The result from the theory states that \itshape the optical transition probability \normalfont demonstrates the power law divergence towards the Fermi energy serving as a cut-off point, and turns out to be enhanced or suppressed depending on the fine details of the transition process. Despite the enormous success in explaining numerous experimental findings on e.g. X-ray absorption spectra features in metals and optical properties of degenerate semiconductors, the underlying physics of the phenomenon seems to be hidden in the truly sophisticated theoretical framework. This is generally repeated that the hole is better scattered by electrons near Fermi surface because they have an ability to scatter into closely located empty states, and the phenomenon is due to the Coulomb interaction. Looking into the details of theory ~\cite{Mahanbook} one can conclude that the origin of the enhancement lies in the increased overlapping of the electron wavefunctions in the initial and the final state of the optical process in case the final state is close enough to the Fermi surface. This feature of wavefunction overlapping resembles the theoretical steps in accounting the excitonic effects in optical spectra of undoped semiconductors ~\cite{Cardonabook} and therefore justifies the use of the expression 'Mahan exciton'.

On the other hand, Friedel oscillations arise as a fine feature of the screening process when the static potential of some defect (typically, a point charge) is present in a degenerate electron system. Instead of electron density monotonically changing apart from the point charge, it reveals oscillations in the form of $\delta n \sim cos(2k_Fr)$, where $r$ is the distance from the defect. The effect does not require any interactions between electrons, and can be viewed as a result of a 'finite size' of electrons with momentum $k_F$ that are most involved in the screening process - again due to the reasoning of free- and occupied electron states coexistence ~\cite{Friedel_naive}, ~\cite{Note_about_momentum_around_defect}.

We are now going to qualitatively describe how Friedel oscillations can directly result to an increased wavefunctions overlapping of the valence band hole and electrons close to the Fermi surface - and finally to a Mahan enhancement. The model system will be a prototypical heterostructure containing a two-dimensional electron system and a neutral acceptor at some distance from the electron plane - the object of our recent experimental study ~\cite{PRB_mass} where recombination enhancement with electrons near $k_F$ was detected for low electron densities, and shown to be suppressed for higher electron densities.

The existence of a the neutral acceptor near the conducting plane of electrons will result to a polarization of the acceptor and the screening of its dipole-like potential by 2D electrons. This is a self-consistent solution of the screening problem and its (most important for the present discussion) consequence will be an appearance of 2D Friedel oscillations in an electron density distribution around the acceptor site ~\cite{JETP_dipole}. The acceptor hole now interacts with two potentials: the unmodified Coulomb attraction to the acceptor core and the varying electron density in 2D plane. Without Friedel oscillations, this would simply give us the usual screening (let's suppose the hole remains in a localized state). However, the  $\sim cos(2k_Fr)$ spatial electron density oscillations will impose the same period oscillations on the hole density. As the density equals to the squared hole wavefunction, this might result, at least for a rather high modulation depth in the hole density distribution, to the half-period oscillations in the hole wavefunction, i.e. it will get a $\sim cos(k_Fr)$ modulation. It is pretty straightforward to anticipate the 'Mahan enhancement' from this point: the Fourier-transformed hole wavefunction (alternatively, the wavefunction in k-space representation) will have increased weight around $k_F$ and thus its overlapping with electrons of that momentum will increase. 

The rigorous theoretical treatment of the suggested problem looks to be a challenging task as it hardly reduces to an effective three-body one, and in fact corresponds to a complicated many-body case if the Coulomb repulsion between electrons is taken into account. It is beyond the scope of the present communication which - we hope - will stimulate the theoretical community to consider the problem in detail.

However, it is instructive to speculate how the limiting case of the free valence hole can be approached from the presented model, and whether the 'Mahan enhancement' phenomenon can be expected in this case. We'd like to proceed in a rather unusual - and prohibited for real physical objects - way: by a gradual reducing of the acceptor core charge value. Once this procedure is started, the 'defect' potential seen by the electron sea changes its character from the dipole-like to the point-charge-like one. More important, in some sense it gets 'stronger' and the screening response of electrons also increases, along with an enhancement of Friedel oscillations depth and the resulting ehnahced modulation of the hole wavefunction. It gravitates towards the form of a plane wave (more correctly, a spherical wave) with momentum $k_F$, and that again promotes the optical processes with electrons of the same momentum. The only driving force for this wavefunction transformation is the Coulomb attraction to the electron density wave - though we never mentioned the specific values for both. It appears that free valence hole will always reconstruct the wavefunction to reach the $k_F$ momentum, even for arbitrary weak Coulomb attraction - unless we recall about the kinetic energy term. So far we did not concern about this contribution, implicitly suggesting its infinite mass. It's clear that the wavefunction transformation costs the $\hbar^2(k_F)^2/2m_h$ addition to the hole total energy, if we assume the hole dispersion to be parabolic with a mass $m_h$. Therefore the phenomenon of 'Mahan enhancement' will be suppressed for large enough $k_F$ values, or, alternatively, for high electron densities. Further theoretical efforts are anticipated in the field.

There are two valuable notes that should be added: 1) the same relation of Friedel-Mahan phenomena can be discussed for the reverse optical process - the light absorption into the final state of neutral acceptor, when the initial state of the charged acceptor is screened by electrons revealing Friedel features again; 2) the X-ray absorption process in metals, when no defect that needs to be screened is present in the initial state, actually leads to the final state when the atomic core needs to be screened - and we again could apply similar reasoning.

To summarize, a conjecture about the intimate relationship between the Friedel oscillations and Mahan exciton(Fermi edge singularity) phenomena is formulated. We qualitatively show how an arbitrary weak Couloumb attraction of the mobile hole to a degenerate electron system can lead to the 'hole acceleration' up to the Fermi-momentum $k_F$, in case the hole mass is infinitely large. The suppression of the 'Mahan enhancement' due to an increase in the hole kinetic energy is explicitly shown, the effect being pronounced as the electron density increases.

\begin{acknowledgments}

We thank A.V. Chaplik for stimulating discussions, and the overall fruitful atmosphere of the XIII Russian Conference on Semiconductor Physics (Ekaterinburg, 2017) where the presented conjecture was conceived. 

\end{acknowledgments}


\begin{thebibliography}{99}
\bibitem{Note_about_momentum_around_defect}

Strictly speaking, momentum is not a good quantum number for electrons in case there is a static defect present in an electron system (it breaks the translational symmetry), and still $k_F$ enters all relevant formulas.
\end{thebibliography}
\end{document}